\documentclass[aps,prl,preprint,superscriptaddress,showpacs]{revtex4-1}
\usepackage{graphicx}

\begin{document}


\title{Measurements of the gate tuned superfluid density in superconducting LaAlO$_3$/SrTiO$_3$}


\author{Julie A. Bert}
\affiliation{Stanford Institute for Materials and Energy Sciences, SLAC National Accelerator Laboratory, 2575 Sand Hill Road, Menlo Park, California 94025, USA}
\affiliation{Department of Physics, Stanford University, Stanford, California 94305, USA}
\author{Katja C. Nowack}
\affiliation{Department of Applied Physics, Stanford University, Stanford, California 94305, USA}
\author{Beena Kalisky}
\affiliation{Department of Applied Physics, Stanford University, Stanford, California 94305, USA}
\affiliation{Department of Physics, Nano-magnetism Research Center, Institute of Nanotechnology and Advanced Materials, Bar-Ilan University, Ramat-Gan 52900, Israel}
\author{Hilary Noad}
\affiliation{Department of Applied Physics, Stanford University, Stanford, California 94305, USA}
\author{John R. Kirtley}
\affiliation{Department of Applied Physics, Stanford University, Stanford, California 94305, USA}
\author{Chris Bell}
\affiliation{Stanford Institute for Materials and Energy Sciences, SLAC National Accelerator Laboratory, 2575 Sand Hill Road, Menlo Park, California 94025, USA}
\author{Hiroki K. Sato}
\affiliation{Stanford Institute for Materials and Energy Sciences, SLAC National Accelerator Laboratory, 2575 Sand Hill Road, Menlo Park, California 94025, USA}
\author{Masayuki Hosoda}
\affiliation{Stanford Institute for Materials and Energy Sciences, SLAC National Accelerator Laboratory, 2575 Sand Hill Road, Menlo Park, California 94025, USA}
\author{Yasayuki Hikita}
\affiliation{Stanford Institute for Materials and Energy Sciences, SLAC National Accelerator Laboratory, 2575 Sand Hill Road, Menlo Park, California 94025, USA}
\author{Harold Y. Hwang}
\affiliation{Stanford Institute for Materials and Energy Sciences, SLAC National Accelerator Laboratory, 2575 Sand Hill Road, Menlo Park, California 94025, USA}
\affiliation{Department of Applied Physics, Stanford University, Stanford, California 94305, USA}
\author{Kathryn A. Moler}
\email{kmoler@stanford.edu}
\affiliation{Stanford Institute for Materials and Energy Sciences, SLAC National Accelerator Laboratory, 2575 Sand Hill Road, Menlo Park, California 94025, USA}
\affiliation{Department of Physics, Stanford University, Stanford, California 94305, USA}
\affiliation{Department of Applied Physics, Stanford University, Stanford, California 94305, USA}

\date{\today}

\begin{abstract}
The interface between the insulating oxides LaAlO$_3$ and SrTiO$_3$ exhibits a 
superconducting two-dimensional electron system that can be modulated by a gate voltage.  
While gating of the conductivity has been probed extensively and gating of the 
superconducting critical temperature has been demonstrated, the 
question whether, and if so how, 
the gate tunes the superfluid
density and superconducting order parameter is unanswered.
We present local magnetic susceptibility, related to the superfluid density, 
as a function of temperature, gate voltage and location.  We show that the
temperature dependence of the superfluid density at different 
gate voltages collapse to a single curve characteristic of a full superconducting gap.
Further, we show that the dipole moments observed 
in this system are not modulated by the gate voltage.
\end{abstract}
\maketitle

Electric field control of conducting channels has allowed 
great innovation in traditional semiconductor 
devices \cite{ahn_electrostatic_2006}. Now heterointerfaces in a 
new class of materials, the complex oxides, have generated significant 
interest because of their gate tunable properties. Specifically, the 
conducting interface formed between the band insulators lanthanum 
aluminate and TiO$_2$ terminated ${100}$ strontium titanate 
(LAO/STO) \cite{ohtomo_high-mobility_2004} exhibits many fascinating 
properties \cite{schlom_oxide_2011} suggesting that an electronic 
reconstruction triggered by the polar/non-polar interface plays an 
important role in the inducing the conductivity in the 
STO \cite{nakagawa_why_2006}. At low temperatures this interface 
displays two-dimensional superconductivity \cite{reyren_superconducting_2007}. 
Additionally, the high dielectric constant of STO at low 
temperatures \cite{sakudo_dielectric_1971} makes applying an electric 
field with a back gate especially effective to tune the 
properties of this superconducting state. 

Caviglia $et\,al.$ showed that with increasing gate voltage, 
$V_g$, the superconducting critical temperature, $T_c$, displayed a 
dome structure and concurrently the normal state resistance 
monotonically decreased \cite{caviglia_electric_2008}. Later work showed 
that the electron mobility and carrier density both increased continuously 
with $V_g$, with the former dominating the $V_g$ dependence of the 
conductivity \cite{bell_dominant_2009}. The evolution of a non-linearity 
in the Hall resistivity as a function of 
$V_g$ \cite{bell_dominant_2009, ben_shalom_tuning_2010} has been 
interpreted by Joshua $et\,al.$ as evidence of electrons populating conduction 
bands with different mobilities \cite{joshua_universal_2011}, implying 
that the ratio of high and low mobility electrons may be tuned by gating. 

Notably, the interface breaks spatial inversion symmetry, opening the 
possibility for spin orbit coupling to impact the electronic properties 
of the interface gas. Two groups reported tuning of the Rashba spin orbit 
coupling (RSOC) inferred from 
magnetoresistance \cite{caviglia_tunable_2010, fete_rashba_2012} and 
measurements of the in-plane critical fields \cite{ben_shalom_tuning_2010}.  
They found opposite dependencies for tuning the strength of the spin 
orbit coupling with $V_g$, making the impact of $V_g$ on the spin 
orbit coupling unclear, possibly suggesting a peak in the spin orbit coupling.

Moreover, the discovery of magnetic patches coexistent with 
superconductivity \cite{bert_direct_2011, li_coexistence_2011, dikin_coexistence_2011} 
and the presence of RSOC originating from the noncentrosymmetric 
nature of the interface have raised 
the possibility of an unconventional superconducting pairing mechanism 
or order parameter 
\cite{michaeli_superconducting_2012,loder_superconductivity_2012,liu_local_2010}. 
However, all previous measurements studying how gating effects the 
properties of the interface used electronic transport, which gives 
limited information about the superconducting state. In this Letter, 
we use local magnetic susceptibility to make the first direct
measurements of the superfluid density in LAO/STO 
and address the question of how the superconducting state evolves with $V_g$.
      
      Measurements were made on a sample with five unit cells of LAO grown at 
$800^{\circ}\,$C and $1.3\times10^{-5}\,$mbar oxygen partial pressure on a TiO$_2$ 
terminated STO substrate.  The growth was followed by a high pressure oxygen anneal, 
$600^{\circ}\,$C in $0.4\,$bar.  The sample was silver epoxied to a piece of copper tape, which 
served as a back gate.  $V_g$ was applied between the copper tape and
the interface, which was contacted by aluminum wirebonds.   Magnetization and susceptibility 
measurements were made using a scanning SQUID (Superconducting Quantum Interference 
Device) \cite{bjornsson_scanning_2001},
with a $3\,\mu$m diameter pick-up loop and a 
concentric field coil for applying a local AC magnetic field.  The pick-up loop is 
sensitive to both the DC static flux and the AC flux resulting from diamagnetic screening 
currents cancelling 
the field from the field coil.  This setup enables simultaneous measurements of 
ferromagnetism and superconductivity in the sample \cite{bert_direct_2011}.
      
      A superconductor will generate screening currents to  
screen an applied field.   The currents extend into a bulk superconductor 
by the penetration depth, $\lambda$. The temperature dependence of $\lambda$
is a probe of the superconducting state.  For a thin 
superconductor of thickness $d$, the 
screening distance is given by the Pearl length 
$\Lambda=2\lambda^2/d$ \cite{pearl_current_1964}.  Using 
a model by Kogan \cite{kogan_meissner_2003}, we extract $\Lambda$ from measurements 
of the screening currents as a function of the distance between the sensor and  the sample.  
$\Lambda$ is related to the superfluid density, $n_s = 
2m^*/\mu_0e^2\Lambda$, where $e$ is the elementary charge, $\mu_0$ the permeability of 
free space, and  $m^*=1.46\,m_e$ the effective electron 
mass measured by \cite{caviglia_two_dimensional_2010} from Shubnikov de Haas on LAO/STO 
interfaces. We repeat these measurements at 
multiple temperatures and gate voltages to map out the superconducting 
state, Fig. 1.  We define $T_c$ as the temperature at which the 
diamagnetic screening drops below our noise level of $0.01\,\Phi_0/$A, 
corresponding to
a minimum detectable $n_s$ of $4-14\times10^{10}\,$cm$^{-2}$.   
The statistical errors 
were smaller than the systematic errors, outlined in gray in Fig. 2a, 
from imprecise knowledge 
of our measurement geometry \footnote{See Supplementary Material for a discussion of
our systematic errors.}.
The systematic errors are fixed for a single
cooldown and represent an overall scaling of $n_s$ which would be the same
for every measurement.  

\begin{figure}[h]
\includegraphics[width = 3.3in]{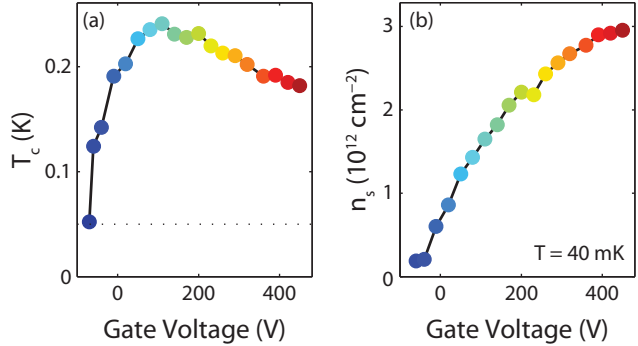}
\label{fig1}
\caption{
a) The critical temperature as a function of gate voltage forms a dome.  
The dashed line represents our lowest measurement temperature. 
b) The superfluid density at our lowest temperature as a function of gate voltage.  
The superfluid density increases monotonically throughout the dome.  
The color scale represents gate voltage and is repeated in Fig. 2.
}
\end{figure}
      
      $T_c$ vs $V_g$ (Fig. 1a) has a maximum $T_c=240\,$mK.   
In the range of applied $V_g$ 
superconductivity can only be eliminated on the underdoped side of the dome,
and $n_s$ grows monotonically with $V_g$, with
$n_s = 3.0\times10^{12}\,$cm$^{-2}$ at the largest $V_g$. (Fig. 1b)   
The carrier density and mobility were measured in a separate cooldown with
no backgate. At $2\,$K the mobility
was $1.02\times10^3\,$cm$^2$/Vs
and the density was $2.05\times10^{13}\,$cm$^{-2}$,
ten times larger than the largest $n_s$ we observed.  
      
      A small ratio of the superfluid density to the normal density is expected 
in the dirty limit, in which the elastic 
scattering time, $\tau$,  
much shorter than the superconducting gap, $\Delta_0$ ($ \hbar/\tau\gg\Delta_0$). $\hbar$
is reduced Plank's constant.  Above $T_c$ 
the normal density of electrons $n$ is given by the optical sum rule 
$n\propto\int_0^{\infty}\sigma_1(\omega)d\omega$, where $\sigma_1$ is the real part of
the conductivity and $\omega$ the frequency. For a metal $\sigma$ is 
sharply peaked near zero frequency, so scattering 
moves spectral weight to higher frequencies.  Below $T_c$, 
a gap opens at $\omega = 2\Delta_0/\hbar$ and the spectral weight 
within that gap collapses to a delta function at the 
origin whose amplitude is proportional to $n_s$ \cite{homes_sum_2004}. Therefore in 
the dirty limit, only a fraction of carriers enter the superconducting state, 
$n_s/n=2\Delta_0/(\hbar/\tau)$.  Using the gate tuned mobilities reported by Bell $et\,al.$, 
$100-1000\,$cm$^2$/Vs \cite{bell_dominant_2009}, 
we expect the ratio $n_s/n$ to be $0.01-0.1$, 
consistent with our measured $n_s$.
\begin{figure}
\includegraphics[width= 3.0in]{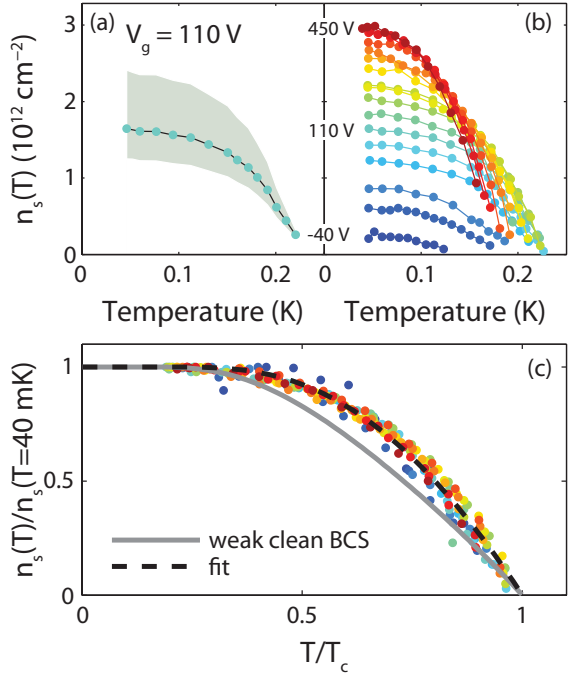}
\label{fig2}
\caption{
a) Superfluid density vs. temperature for $V_g=110 V$, 
the peak of the superconducting dome.  The gray area shows systematic error. 
b) Superfluid density vs. temperature for 
every gate voltage.  The colors represent the same $V_g$ from Fig. 1. 
c) Normalized curves from b).  
The gray line shows the temperature dependence of a weakly interacting clean 
BCS s-wave superconductor ($\Delta = 1.76$ and $a=1$).  The black dashed line is 
a fit to the data ($\Delta = 2.2$ and $a=1.4$).
}
\end{figure}      
      
      We now look at the temperature dependence of the superfluid density.  Fig. 2b plots 
$n_s$ vs. $T$ for all $V_g$ across the dome.
Strikingly, when normalizing the curves they collapse (See Fig. 2c), 
showing that within our experimental errors there is no change in the superconducting gap 
structure with electrostatic doping.  Furthermore, the collapse is reproducible over multiple 
positions, sweeps of $V_g$, and samples \footnote{Similar behavior was seen in a 
separate 10 uc sample.}.
      
      The temperature dependence of the superfluid density is a direct probe of the 
superconducting order parameter. It can be used to distinguish BCS superconductors from 
unconventional superconductivity.  We fit the normalized curves to a phenomenological BCS 
model with two parameters $\Delta$ and $a$ \cite{prozorov_magnetic_2006}.   $\Delta$ 
scales the superconducting gap $\Delta_0 = \Delta{k_BT_c}$.  $a$ is a shape parameter that 
determines how rapidly the gap opens below $T_c$, $n_s \propto 1-(T/Tc)^{2a}$ \footnote{
See supplementary material for the details of the phenomenological model}.  
$\Delta=1.76$ and $a=1$ for an clean s-wave BCS superconductor with 
weak coupling \cite{prozorov_magnetic_2006}, plotted as the gray line in Fig. 2c.
The fit to our data gives $\Delta = 2.2$ and $a=1.4$. This is consistent with a BCS
description with increased coupling or disorder.  Both will theoretically increase 
the gap and the $a$ parameter \cite{carbotte_properties_1990}, shifting the curve
up and to the right.

    The flattening at low temperature indicates 
fully gapped behavior with a gap that is larger than BCS weak-coupling s-wave.  
Our lowest 
measurement temperature is $~1/6$ of $T_c^{\rm{max}}$, and 
$n_s$ remains flat (within 3\%) up to 35\% of $T_c$.  A full gap indicates 
the absence of low energy 
quasiparticle excitations, ruling out order parameters with nodes in the Fermi surface.  
Furthermore, the steep rise of $n_s$ near $T_c$ and the 
absence of a kink in the functional form rule out most weak coupling two band 
models \cite{prozorov_london_2011}, because a second smaller gap will slow 
the onset of superconductivity near $T_c$.  Two gaps of similar 
size, both larger than the BCS gap or a dominant single large gap with 
second smaller amplitude gap, could reproduce the data.
\begin{figure}
\includegraphics[width=3.0in]{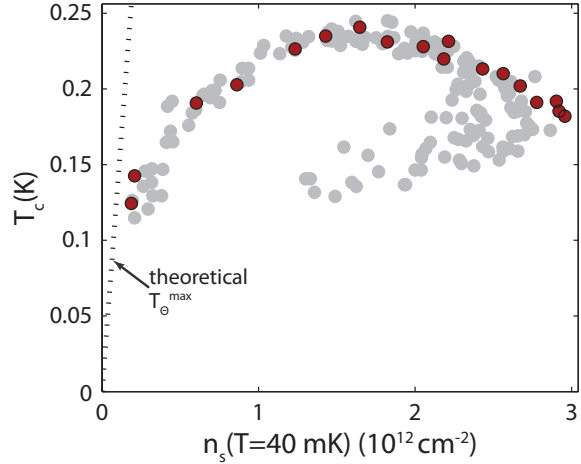}
\label{fig3}
\caption{
Critical temperature vs. the superfluid density at lowest 
temperature ($T\sim40\,$mK).  The red points are the data from Figure 1 
and the gray dots represent additional data sets.  The 
dotted line is the theoretical
phase fluctuation temperature from ref \cite{emery_importance_1995}, 
which may be limiting 
the critical temperature on the underdoped side of the dome.
The bimodal distribution on the overdoped side is due to 
inhomogeneity that locally suppresses $n_s$ in different regions of the
sample while the $T_c$ remains the same. See also Fig. 4. 
}
\end{figure}      

The low $n_s$ in the underdoped region  
may result in suppression of $T_c$ by thermal phase fluctuations.  Such fluctuations 
would result in a linear temperature dependence of $n_s$ in the underdoped region.  
Following reference  \cite{emery_importance_1995}, we calculate a phase ordering temperature, 
$T^{max}_{\theta}= A\hbar^2n_s(0)/4m^*$, where $A=0.9$ in two dimensional 
systems. Fig. 3 shows $T_c$ vs $n_s(40\,\rm{mK})$, 
additionally $T^{max}_{\theta}$ is plotted
as a linear function of $n_s$: the line does not suggest a fit to our data.  
We have insufficient data at
the lowest superfluid densities to make any statement about the functional form of 
$T_c(n_s)$ in the region where phase fluctuations may be limiting $T_c$.
Nevertheless, the proximity of the phase 
ordering line to the underdoped data suggests that phase fluctuations may 
drive the abrupt decrease of $T_c$.  

Given the 2D nature of the  
superconducting system we expect a BKT transition, where unbinding of vortex 
anti-vortex pairs suppresses superconductivity and results 
in a discontinuous jump in $n_s$ near $T_c$.  The jump should occur at finite 
superfluid density $n_{s}=2m^{*}T_c/\pi\hbar^2$ \cite{pokrovskii_magnitude_1988}.
For the maximum $T_c = 240\,$mK a BKT transition should occur at 
$5\times10^{10}\,$cm$^{-2}$, which is too close to our measurement threshold to establish a BKT 
jump in our $n_s$ vs. $T$ curves.

Are our observations consistent with a simple $s-$wave 
order parameter from doped STO \cite{koonce_superconducting_1967} or a two gap mixed 
state induced by symmetry breaking at the interface?  Rashba spin orbit 
coupling (RSOC), induced by the structural inversion 
asymmetry, is expected to lift the spin degeneracy 
and split the energy bands \cite{winkler_spin-orbit_2003}.
Additionally, RSOC breaks parity and consequently mixes singlet 
and triplet states resulting in an $s-$wave component $\Delta_s$ mixed with a triplet induced 
d-vector $\mathbf{d(k)}=\hat{x}k_y-\hat{y}k_x$ \cite{frigeri_superconductivity_2004, 
liu_local_2010}.  Mixing results in two gaps, $\Delta=\Delta_s\pm|d_k|$, whose 
magnitudes depend on the weights of the singlet and triplet components.  Varying 
the relative weights changes the density of states, but always results in two 
fully gapped Fermi surfaces except for the special case where the s-wave singlet and triplet 
gaps are the same and accidental line nodes form on one band \cite{liu_local_2010}.
      
      Other reports \cite{caviglia_tunable_2010, ben_shalom_tuning_2010} 
have demonstrated significant tuning of the strength of RSOC
with $V_g$. An open question, of particular importance to 
testing this two gap picture, is how do 
the weight of the two components change with $V_g$.  Our results, showing a 
consistent functional form for $n_s$ vs. $T$ across all 
$V_g$, show that the superconducting gap structure does not change with  
$V_g$ consequently the relative gap weights do not change with $V_g$.  
The effect of RSOC on the band structure may depend on the 
chemical potential which is also tuned by the gate.   Therefore the insensitivity of 
superconductivity to $V_g$ cannot completely rule out a RSOC induced two gap scenario.  Yet, our 
second observation of the fast opening of the gap near $T_c$ and the compatibility of the data 
with a single gap BCS model limits two gap models.  Both gaps must be 
larger than the BCS s-wave gap to capture both the fast rise and flat low temperature 
dependence of the data \footnote{See supplementary materials for a discussion of multiple gaps.}.  

      Finally, disorder may play a role in washing out the triplet 
component.  As stated above, the LAO/STO system is a dirty superconductor, with 
$\hbar/\tau>>\Delta$.  Disorder averaging has very little impact on the isotropic s-wave 
component but may eliminate the triplet component.  

     In short, our data is most consistent 
with a single gap.  We cannot rule out the presence of two gaps, but our 
observations limit their size and $V_g$ dependence.
\begin{figure}[ht!]
\includegraphics[width=3in]{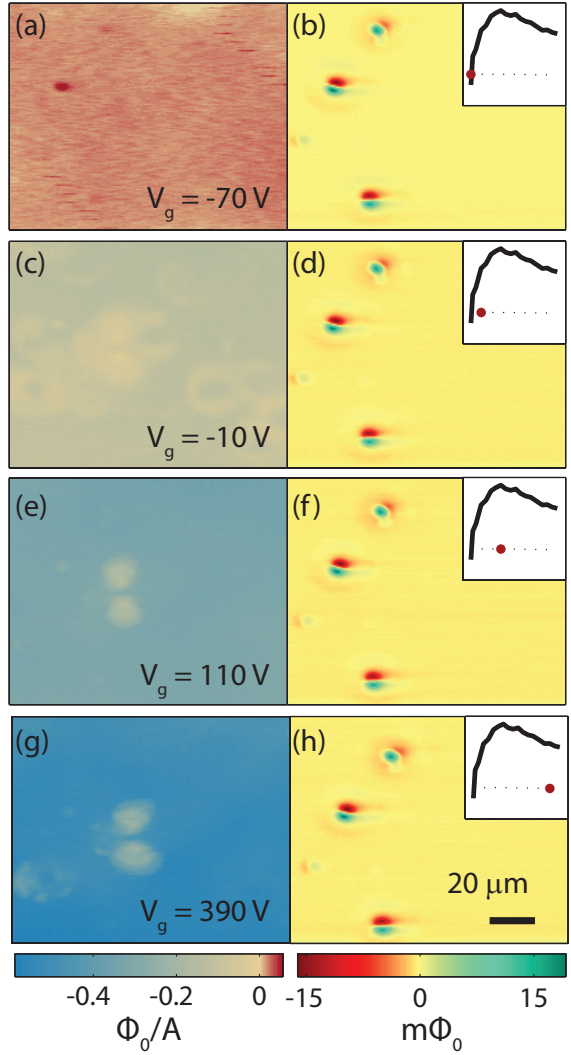}
\label{fig4}
\caption{
Susceptometry (left) and magnetometry (right) 
at $80\,$mK at different gate voltages. (Inset) Reproduction of the
$T_c$ dome from FIG 1 showing the relative location of $V_g$ in each panel.
a-b) The sample is no longer superconducting and has a paramagnetic 
response. Individual ferromagnetic dipoles are also visible in the paramagnetic image.  
c-d) Superconductivity appears and the landscape is relatively inhomogeneous.  
e-f) Peak of the superconducting dome, most inhomogeneity disappears. 
g-h) Excess inhomogeneity returns on the overdoped side of the dome.  
The ferromagnetic patches do not 
change with $V_g$ and remain when superconductivity is gone.
}
\end{figure}      
      
     Our scanning SQUID system allows two dimensional mapping of superconductivity 
and magnetism at different $V_g$. Fig. 4 shows simultaneously imaged susceptometry 
and magnetometry scans of the same region at $80\,$mK for four different $V_g$.  The 
inhomogeneity in the diamagnetic screening is very large in the underdoped region 
($V_g = -10\,$V) and re-enters the image in the overdoped region ($V_g= 
390\,$V).   The least inhomogeneity is observed at optimal doping, although it does not 
disappear.   
In contrast the ferromagnetic patches are insensitive to $V_g$ with a constant 
magnitude and orientation for all $V_g$. This behaviour was also observed on $15, 10\,$ and
$3.3\,$ uc samples, showing the electron 
density that is modified by $V_g$ does not appear to influence the ferromagnetism.
      
      In conclusion, we presented the first measurements of 
the superfluid density as a function of temperature at 
multiple gate voltages throughout the superconducting dome in LAO/STO heterostructures.    
The temperature dependence of $n_s$ is well described by a fully gapped BCS model. Moreover, 
the normalized $n_s$ vs. $T$ curves collapse to a single functional 
form indicating there is no change in the gap structure with $V_g$. Although we cannot
rule out a two gap mixed singlet/triplet model, the insensitivity of the superconducting
state to $V_g$ and the large slope near $T_c$ limit two gap scenarios.
Specifically, both gaps must be larger than the BCS s-wave gap and their relative size 
cannot change throughout the dome.  
A future experiment to distinguish between these two scenarios may be to gate the 
superconductivity in the presence of an in-plane field, which can change the relative magnitude 
of triplet and singlet gaps.  Alternatively, samples in the clean limit may reveal a clearer
two gap structure.  Additionally, we found that the magnitude and orientation of the 
ferromagnetic patches that coexist with superconductivity are unchanged by $V_g$, 
while at the same time $n_s$ goes from zero to 
$3.0\times10^{12}\,$cm$^{-2}$.  This shows the population of electrons that is modified 
by the gate is separate from the electrons that contribute to the ferromagnetic order.    

\begin{acknowledgments}
We thank S. A. Kivelson, E. A. Kim, M. H. Fischer, 
S. Raghu, A. Kampf and I. Sochnikov for useful discussions and
M. E. Huber for assistance in SQUID design and fabrication. 
Work was supported by the US Department of Energy, 
Office of Basic Energy Sciences, Materials Sciences 
and Engineering Division, under award DE-AC02-76SF00515. 
B.K. acknowledges support from FENA.
H.N. acknowledges support from Stanford Graduate Fellowship.
K.C.N acknowledges support from NSF Grant Nos. DMR-0803974.
\end{acknowledgments}

%

\newpage
\noindent
\textbf{\large{SOM: Gate tuned superfluid density measurements of superconducting LaAlO$_3$/SrTiO$_3$}}
\normalsize
\section{Discussion of Systematic Errors}
The accuracy of our superfluid density measurement is dominated by systematic errors which arise from 
insufficient knowledge of the physical parameters of our SQUID sensor and piezoelectric scanner.  Our 
measurement of the superfluid density, $n_{s}$, relies on extracting the Pearl length, $\Lambda$, from 
fits to approach curves.  
\begin{equation}
n_s = \frac{2m^*}{\mu_0e^2\Lambda}
\end{equation}
An approach curve measures the diamagnetic susceptibility as a function of the sensor height above the 
sample.  Our SQUID sensor consists of a pair of concentric current carrying wires called the 
field coil and pick-up loop.   

\begin{figure}[h]
\centering
\includegraphics[width=3.3in]{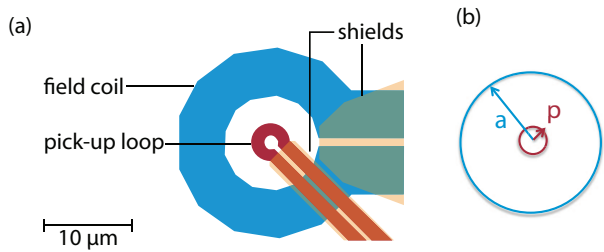}
\label{som_fig1}
\caption[Schematic of SQUID tip]{
a) Actual layout of the SQUID field coil, pick-up loop and shields. b) Approximations
to the actual layout used by Kogan \cite{kogan_meissner_2003}.
}
\end{figure}

We follow a model developed by Kogan which treats the SQUID's field coil as a circular 
current loop of radius, $a$ \cite{kogan_meissner_2003}.  When the loop is brought near a 
superconducting thin film, the Meissner response of the film detected by the pick-up loop can be 
expressed as
\begin{equation}
\Phi(h)=\mu_0\pi{ap}\int_0^{\infty}dk\frac{1}{1+\Lambda{k}}e^{-2kh}J_1(ka)J_1(kp),
\label{eq:kogan}
\end{equation}
where $p$ is the pick-up loop radius, and $\mu_0$ is the magnetic 
constant.  This integral gives a value for the diamagnetic susceptibility, 
$\Phi$, at a height $h$ above the 
sample.  Six physical parameters enter equation (\ref{eq:kogan}): the radius of 
the pick-up loop $p$, the 
radius of the field coil $a$, the piezo calibration from volts to microns $V_c$, 
the distance between the 
pick-up loop and the sample when the SQUID makes contact $h_0$, the offset of the susceptibility far 
from the sample $\Phi_{off}$, and a background slope $m$.  We convert the voltage applied to the 
z-bender, $V_z$, to a height $h = V_cV_z + h_0$.  The susceptibility seen by the SQUID is
\begin{equation}
\Phi_{SQ} = \Phi+\Phi_{off}+mh. 
\end{equation}
 Consequently, our fits for $\Lambda$ depend on the accuracy of our knowledge of the other 
parameters.

We start with estimates of $p$ and $a$. We can make accurate measurements of the two radii using an 
optical microscope; however these wire loops have a finite width and leads that deform their magnetic 
response with respect to the perfectly circular loops in Kogan's model.   Using numerical methods, we 
calculate the source field using the measured dimensions of our non-ideal coils.   We find the non-ideal 
nature of the field coil and pick-up loop results in a 15\% error on the fitted value of $\Lambda$.  
This error works out to few hundred microns on our shortest $\Lambda$ fits.

We now address the errors associated with our bender constant $V_c$ and height offset $h_0$.  We 
don't have accurate calibrations for these parameters, but we do know that 
these values should be the same for 
every touchdown curve.  We fit hundreds of approach curves using $\Lambda$, $V_c$, and $h_0$ as 
free parameters, and assembled histograms of the fitted $V_c$ and $h_0$ values.  From the histograms 
we were able to extract a best value and variance, $\sigma$.  We then use the error propagation 
equation to relate the variances in $V_c$ and $h_0$ to an error in $\Lambda$.
\begin{equation}
\sigma_{\Lambda}^2\simeq\sigma_{V_c}^2\left(\frac{\partial{\Lambda}}{\partial{V_c}}\right)^2+\sigma
_{h_0}^2\left(\frac{\partial{\Lambda}}{\partial{h_0}}\right)^2+...+2\sigma_{V_ch_0}^2\left(\frac{\partial
{\Lambda}}{\partial{V_c}}\right)\left(\frac{\partial{\Lambda}}{\partial{h_0}}\right)+...
\end{equation}
 The propagation equation yielded an error of about 1 mm on our shortest $\Lambda$ fits.
This is a systematic error and is the same for every touchdown curve in the cooldown.  It may 
change the overall calibration for $n_s$, but it will not change the trends in $n_s$ vs $V_g$ or
$n_s$ vs $T$.    

We added the systematic errors from the sensor coils, bender calibration and height offset.  The total 
systematic error is show as the gray outline shown in Fig. 2a of the main text.  
The error from the bender and offset 
dominates the error from the non-ideal nature of the pick-up loop and field coil. 
\section{Discussion of Phenomenological BCS Fits}
We compare our normalized plots of superfluid density vs. temperature to a phenomenological BCS 
model with an isotropic s-wave superconducting gap. The normalized superfluid density, 
$n_s/n_s(T=0)$, was given by 
Prozorov and Giannetta \cite{prozorov_magnetic_2006}
\begin{equation}
\frac{n_s}{n_s(T=0)}=1-\frac{1}{2T}\int_0^{\infty}\cosh^{-
2}\left(\frac{\sqrt{\epsilon^2+\Delta^2(T)}}{2T}\right)d\epsilon,
\label{eq:rho}
\end{equation}
where $T$ is the temperature and $\Delta(T)$ is the superconducting gap function.  The gap can be 
written \cite{gross_anomalous_1986} as 
\begin{equation}
\Delta_0(T)=\Delta_0(0)\tanh\left(\frac{\pi{T_c}}{\Delta_0(0)}\sqrt{a\left(\frac{T_c}{T}-1\right)}\right).
\label{eq:gap}
\end{equation}
$\Delta_0(0)$ is the zero temperature energy gap and $a$ is a shape parameter which determines how 
fast the gap opens.  Near the critical temperature the superfluid density can be approximated as 
$n_s=1-(T/T_c)^{2a} $.  For an isotropic s-wave gap $\Delta_0(0)=1.76k_BT_c$ and $a=1$.
We use equations (\ref{eq:rho}) and (\ref{eq:gap}) to fit our data and find $\Delta_0(0)=2.2k_BT_c$ and 
$a = 1.4$.  This is the dashed line plotted with the data in Fig. 2c of the main text.
\section{Discussion of Two Gaps in BCS}
We can use equations (\ref{eq:rho}) and (\ref{eq:gap}) to generate a phenomenological two-gap 
expression \cite{luan_local_2011}.
\begin{equation}
n_s(T) = p{n_{s1}}(T)+(1-p){n_{s2}}(T)
\end{equation}
Fig. 6 shows plots of the superfluid density for two gaps of equal weight ($p=.5$) with different 
physical parameters. 
\begin{figure}[h]
\includegraphics[width=0.8\textwidth]{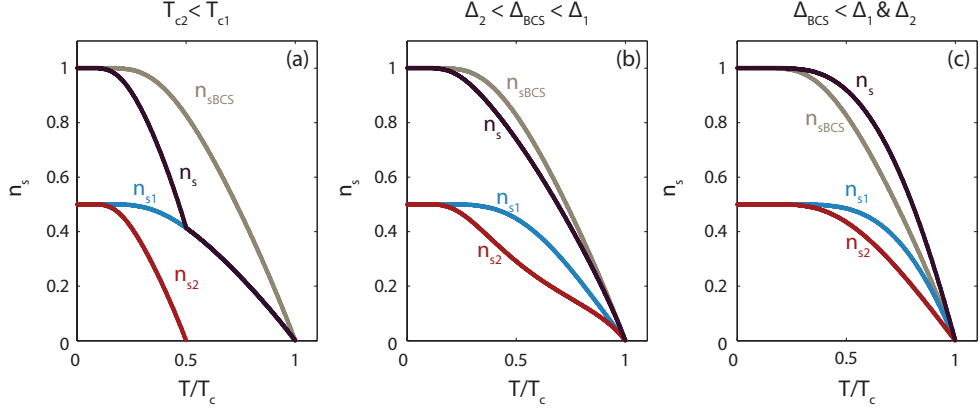}
\label{figSOM1}
\caption{
Comparison of a two-gap superfluid density $n_s$ with a single gap BCS 
superfluid density $n_{s\rm{BCS}}$.  In all three plots $n_{s1}= pn_{s1}$ and $n_s2=pn_{s2}$ with 
a) Plots of two gaps with $\Delta_1=\Delta_2=1.76$ and $a_1=a_2=1$ but two different 
critical temperatures. b) Plots of two gaps with $\Delta_1=2.2$, $\Delta_2=1.1$, and $a_1=a_2=1$.
c) Plots of two gaps with with $\Delta_1=3$, $\Delta_2=2$, $a_1=1.8$ and $a_2=1$.  Only in c) where 
both gaps are larger than $\Delta_{\rm{BCS}}=1.76$ can we generate a total superfluid density 
that opens faster than BCS.  
}
\end{figure}
The only combination that can support a superfluid density function that rises faster than BCS near $Tc$ 
has two gaps that are larger than the BCS gap.

\end{document}